# Complexity Analysis of Approaching Clinical Psychiatry with Predictive Analytics and Neural Networks

Soaad Q. Hossain

*Department of Philosophy, Joint Centre of Bioethics, Department of Computer and Mathematical Sciences, University of Toronto, Canada*

**Abstract:** As the emerging field of predictive analytics in psychiatry generated and continues to generate massive interest overtime with its major promises to positively change and revolutionize clinical psychiatry, health care and medical professionals are greatly looking forward to its integration and application into psychiatry. However, by directly applying predictive analytics to the practice of psychiatry, this could cause detrimental damage to those that use predictive analytics through creating or worsening existing medical issues. In both cases, medical ethics issues arise, and need to be addressed. This paper will use literatures to provide descriptions of selected stages in the treatment of mental disorders and phases in a predictive analytics project, approach mental disorder diagnoses using predictive models that rely on neural networks, analyze the complexities in clinical psychiatry, neural networks and predictive analytics, and conclude with emphasizing and elaborating on limitations and medical ethics issues of applying neural networks and predictive analytics to clinical psychiatry.

## Introduction

Psychiatry is one of the specializations in medicine that is heavily pressured due to the increase in demand for mental health service and the performance of psychiatry when it comes to diagnoses and treatments of mental disorders. The World Health Organization (WHO) reported 322 million cases of depressive disorder around the world and 264 million cases of anxiety disorder (World Health Organization, 2017). Combining only these two types of mental disorder cases results in a total of 586 million cases of mental disorders, a number that is said to increase overtime. Predictive analytics has been presented as an innovative tool promising to revolutionize the practice of psychiatry by enabling psychiatry to effectively and optimally operate on many levels including in diagnosis phase and intervention strategy planning.

While predictive analytics may sound promising and worthwhile, if it is misused or misemployed, then consequences ranging from minor to extremely severe can occur. What makes these consequences worse is that the involvement of a patient's mental health is at stake during such cases. The more harm is caused to a patient mentally, the more challenging it will be treat that patient, and time and resources will be used. With issues such as human and medical errors, biases, data manipulation, and misinterpretation of data and evidence being current existing problems within health care and medicine, even with the capabilities and prior positive performance of predictive analytics in other industries, there is no guarantee that it will perform as expected when it is used in clinical psychiatry. Also, as the application of predictive analytics has never been used in psychiatry before, there is no clear understanding of what kind of lethal consequences could be produced from applying predictive analytics to clinical psychiatry. This implies that there is not much research or testing done pertaining to the application of predictive analytics in psychiatry.

This paper will use literatures to provide descriptions of selected stages in the treatment of mental disorders and phases in a predictive analytics project, approach mental disorder diagnoses using predictive models that rely on the commonly used machine learning algorithm known as neural networks, analyze the complexities in clinical psychiatry, neural networks and predictive analytics, and conclude with emphasizing and elaborating on limitations and medical ethics issues of applying predictive analytics backed by neural networks to both diagnoses and treatments in the practice of psychiatry.

## Method

Through using published literature and open-access internet resources, we identified and reviewed information on psychiatry and predictive analytics. The results were analyzed through combining information and knowledge from clinical psychiatry, neural networks, and predictive analytics. We use that information to analyze the complexity of clinical psychiatry, neural networks, and predictive analytics.

## Results

### Literature Search

The literature search resulted in retrieving research articles, medical manuals, critical analyses, and literature reviews to understand steps in diagnosing mental disorders in psychiatry and phases in predictive analytics, then combining them to create predictive models that use neural networks to diagnose mental disorders.

### Clinical Psychology and Psychiatry

Within the treatment of mental disorders, there is a total of 7 sequential stages that must be taken, which those sequential stages are (Kilgus, Maxmen & Ward, 2009):

1. Case Context and Method
2. Case Description
3. Case Formulation and Treatment Approach
4. Treatment Plan and Goal
5. Relational Factors
6. Course of Treatment and Monitoring of Treatment Progress
7. Treatment Outcome

As the first 3 stages are most relevant to the diagnosis of mental disorders, we will investigate and elaborate on those three stages.

**Case Context and Method Stage Summary:** In this stage, description of the treatment setting, relevant context, and sources of data are documented. For the treatment setting, this varies on a case by case basis. For example, cases involving treatment for substance use disorders is delivered at fluctuating levels of care in many diverse settings, which those setting could be outpatient, intensive outpatient, partial hospitalization, residential, and inpatient treatment (National Institute on Drug Abuse, 2014). For relevant content, content would typically include things like patient's medical history (i.e. allergies and current or most recent substances

used). Data collected in this phase includes and is not limited to:

- Case notes
- Audio sessions
- Videotaped sessions
- Patient self-report
- Clinician or therapist self-report
- Questionnaires
- Brain scan

The *Diagnostic and Statistical Manual of Mental Disorders* (DSM-5) Self-Rated Level 1 Cross-Cutting Symptom Measure – Adult is often used as a self- or informant-rated measure to assess mental health domains that are important across psychiatric diagnoses. The measure is intended to assist clinicians with identifying additional areas of inquiry that may have a significant impact on the patient's treatment and prognosis, and can be used to track changes in the patient's symptom(s) presentation over time.

DSM-5 Self-Rated Level 1 Cross-Cutting Symptom Measure—Adult

**Table 1:** DSM-5 Self-Rated Level 1 Cross-Cutting Symptom Measure - Adult (DSM-5, 2013).

**Case Description Stage Summary:** In this stage, description of the treatment context, assessment of presenting problem(s) and situation, and patient(s) are documented. Essential information covered within this stage would pertain to:

- Explanatory models of illness
- Cultural background (including ethnicity, indigenous heritage, intersecting identities of race, social class, nationality, gender, age, religion, sexual orientation, religion, disability, etc.)
- Dimensions of psychosocial history (including stressors, strengths, supports, family history, and timing of help-seeking

Additionally, medical description of the treatment context, assessment of presenting problem(s) and situation, and client(s) are documented.

Combining the first two stages should allow a clinician to complete the *Mental Status Examination* (MSE), a measure that assist clinicians with performing diagnoses.

### Mental Status Exam

| Client Name | | | Date | | | |
|---|---|---|---|---|---|---|
| **OBSERVATIONS** | | | | | | |
| Appearance | ☐ Neat | ☐ Disheveled | ☐ Inappropriate | ☐ Bizarre | ☐ Other | |
| Speech | ☐ Normal | ☐ Tangential | ☐ Pressured | ☐ Impoverished | ☐ Other | |
| Eye Contact | ☐ Normal | ☐ Intense | ☐ Avoidant | ☐ Other | | |
| Motor Activity | ☐ Normal | ☐ Restless | ☐ Tics | ☐ Slowed | ☐ Other | |
| Affect | ☐ Full | ☐ Constricted | ☐ Flat | ☐ Labile | ☐ Other | |
| Comments: | | | | | | |
| **MOOD** | | | | | | |
| ☐ Euthymic | ☐ Anxious | ☐ Angry | ☐ Depressed | ☐ Euphoric | ☐ Irritable | ☐ Other |
| Comments: | | | | | | |
| **COGNITION** | | | | | | |
| Orientation Impairment | ☐ None | ☐ Place | ☐ Object | ☐ Person | ☐ Time | |
| Memory Impairment | ☐ None | ☐ Short-Term | ☐ Long-Term | ☐ Other | | |
| Attention | ☐ Normal | ☐ Distracted | ☐ Other | | | |
| Comments: | | | | | | |
| **PERCEPTION** | | | | | | |
| Hallucinations | ☐ None | ☐ Auditory | ☐ Visual | ☐ Other | | |
| Other | ☐ None | ☐ Derealization | ☐ Depersonalization | | | |
| Comments: | | | | | | |
| **THOUGHTS** | | | | | | |
| Suicidality | ☐ None | ☐ Ideation | ☐ Plan | ☐ Intent | ☐ Self-Harm | |
| Homicidality | ☐ None | ☐ Aggressive | ☐ Intent | ☐ Plan | | |
| Delusions | ☐ None | ☐ Grandiose | ☐ Paranoid | ☐ Religious | ☐ Other | |
| Comments: | | | | | | |
| **BEHAVIOR** | | | | | | |
| ☐ Cooperative | ☐ Guarded | ☐ Hyperactive | ☐ Agitated | ☐ Paranoid | | |
| ☐ Stereotyped | ☐ Aggressive | ☐ Bizarre | ☐ Withdrawn | ☐ Other | | |
| Comments: | | | | | | |
| **INSIGHT** | ☐ Good | ☐ Fair | ☐ Poor | Comments: | | |
| **JUDGMENT** | ☐ Good | ☐ Fair | ☐ Poor | Comments: | | |

Provided by TherapistAid.com © 2013

**Table 2:** Example of a Mental Status Examination (TherapistAid, 2013).

**Case Formulation and Treatment Approach Stage Summary:** In this stage, medical description of the clinician's or therapist's conceptualization of the case as it guided the treatment approach for the target problem(s). At this stage, information on the relevant theoretical, research and/or sociocultural basis for the approach to treating the target problem(s) are covered. It is important to note that the relevant research would encompass qualitative and quantitative studies on therapeutic and other practice-based evidence such as other case studies.

*Clinical Psychiatric Complexity Analysis*

In analyzing complexities in clinical psychiatry, we did not use the definition of complexity according to medicine, but rather the definition of complexity in computer science. We define complexity as the inherent difficulty in solving problems. There are two kinds of complexities apply to the practice of psychiatry: task complexity and patient complexity (Islam, Weil & Del Fiol, 2016).

**Task Complexity:** Task complexity in clinical psychiatry is very intricate due to the complexity and complexity factors associated with performing tasks within the diagnosis and treatment of mental disorders. Complexity factors include and are not limited to: clarity, quantity, diversity, inaccuracy, redundancy, conflict, pressure, and change. The complexity factors exist and impact all 7 stages of treatments of mental disorders, which the complexities within each stage vary and differ from a case to case basis.

Some clinicians may find significant complexities in the Relational Factors Stage (description between the clinician and the patient) due to the mental disorder of their patient while other clinicians may find significant complexities in the Case Formulation and Treatment Approach Stage due to the conflict of deciding which mental disorder best describes and applies to their patient. What makes complexities from clinical psychiatry different from other medical fields is that the complexity of solving psychiatric problems could not only be difficult in some cases, but they can also be impossible as well. For cases such as borderline personality disorder, the National Institute of Mental Health has mentioned that it has historically been perceived as a disorder that is difficult to treat (NIMH, 2017). The reason for this could be due to the patient's inability to properly maintain relationships, negative impulsive behavior, or even intense anger or depression experienced by the patient. This heavily impacts the task complexity of clinicians in a negative way.

Complexities for this case could range from conflicts between the clinician and patient arising, to difficulty convincing the patient to follow a treatment plan and sticking with the treatment plan. These would then make it challenging for the clinician to complete all 7 sequential stages successfully.

**Patient Complexity:** Patient complexity in medical patients involves complexity factors from multiple dimensions. These dimensions include: physical health, mental health, demographics, social capital, and health and social experiences. The complexity factors found within those dimensions include and are not limited to: psychological distress, psychiatric illness, cognitive impairment, poor social support, self-management challenges, loss of physical functioning, and ethnic disparities (Islam, Weil & Del Fiol, 2016).

Patient complexity in psychiatric patients is interesting in the sense that depending on the mental disorder, it is fairly challenging to determine what dimensions are dominant along with which dimensions impact the complexity itself. The reason for this is because some complexity factors are visible while others are not. Physical disabilities tend to be more noticeable than financial situation of an individual. Social and personal experiences are practically invisible until the patient themselves state them. While clinicians can make assumptions about the experiences that a patient may have gone through, the truth will never be known until the patient confesses about them. However, even then, some complexity factors can contribute to how the patient explains their experience to the clinician. If a patient has went through traumatic experiences that resulted in them having trust issues, then the likelihood that the patient says the truth may not be as high as a clinician might expect it to be. Thus, psychiatric patient complexity can be said to be one of the challenging types of patient complexity in the field of medicine.

*Predictive Analytic Project Overview*

Predictive analytics is an analytical tool that uses data, statistical algorithms and predictive modeling techniques to identify the likelihood of a future event based on historical data. It aims to build predictive models to allow an individual to get a better understanding of what will happen in the future. Predictive analytics is widely used in many industries. For example, predictive analytics is used in banking and financial services to detect and reduce fraudulent activities. Just as predictive analytics helped other industries, professionals believe that it can also help within the field of medicine as well. Within all industries, predictive analytics project follows through the same development phases needed to complete to ensure the validity and applicability of predictive models. Those phases are as followed along with their summaries:

Phase 1: Objectives – Defining Objectives
Phase 2: Data Collection – Acquiring Data
Phase 3: Model Development – Building the Model
Phase 4: Model Application – Using the Model in
　　　　 Practical Settings

**Objectives Summary:** In this phase, description of the desired scope of the predictive model, and whether the prediction of an unknown quantity required would be reported and documented.

**Data Collection Summary:** In this phase, description of how the informative features (predictors) were chosen is documented. It is said that if prior data is lacking, then it would be better to base the predictive models exclusively on easily obtainable data; even if the model's power is slightly decreased (Hahn., Nierenberg & Whitfield-Gabrieli, 2017). Also in general, any data set acquired for group-level investigations may be suitable for predictive model development.

**Model Development Summary:** In this phase, description of the machine learning approach and how the predictive model is generated is documented. In practice, the machine learning approach is empirically determined based on the training data.

**Model Application Summary:** In this phase, description of how a validated predictive model can be deployed, and how can future validity of the predictive model can be ensured is reported and documented.

*Predictive Model Selection*

In understanding the 7 sequential stages needed for treatment of mental disorders, stages of predictive analytics project and neural networks, we found that predictive analytics could be applied to the diagnosis of mental disorders in clinical psychiatry. The predictive model that we hypothesized would be a classification model that would be a neural network-based model. Classification models are models that attempt to draw a conclusion from observed values. The reason for selecting neural networks over other predictive modeling techniques is because they can handle complex nonlinear relationships in data. Additionally, from past applications of neural networks, they have been used in the past on gene identification, genetic interaction, and protein structure prediction (Hu et al., 2016), making it a viable pick for clinical psychiatry.

*Neural Networks Overview*

To get a general sense of how neural networks learns, consider the neural network in figure 1., with $X_1$ and $X_2$ each with weights associated with them, as input nodes in the input layer. The input within the input layer are then passed through the subsequent hidden nodes in the hidden layer, extracting a different set of high-level features, until eventually reaching the output node in the output layer. By the time the output layer is reached, if the requirements for activation are met, then the output node is expected to return some sort of predicted value, else a false predicted value will be returned or an undesired value will be returned.

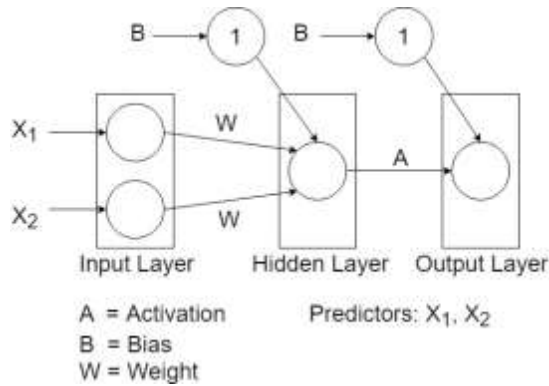

**Figure 1.** Basic neural network model.

The linear equation that would be used would be something like the following:

$$Y = W_1 X_1 + W_2 X_2 - B$$

Which if Y satisfies all the activation requirement, then the expected output would be a predicted value. The values of $X_1$ and $X_2$ are information taken from a training data. The function can be expanded to *n* number of values as well.

$$Y = \sum_{i=0}^{n} W_i X_i - B$$

*Predictive Analytics Project*

To make the clinical psychiatric predictive analytic project for diagnoses, we then did the following:

**Phase 1:** In following the phases required for a predictive analytics project, we first set the desired scope of the predictive model to determine what mental disorder a patient has.

**Phase 2:** For data collection, we would use the patient's static characteristics, which that data collected would the outline of the psychiatric assessment, which contained 8 sections: (1) history, (2) mental status examination, (3) auxiliary data, (4) summary of principle findings, (5) diagnoses, (6) prognosis, (7) biopsychosocial formulation, and (8) plan, and create training data containing the psychiatric data.

**Phase 3:** To develop the predictive model, which for our case we will be using a classification model and training data as we are attempting to draw a conclusion of what mental disorder the patient has from the observed static characteristics of the patient, we used the predictive modeling technique neural networks to create the classification model. The neural network would take all vital signs and laboratory studies from those sections as predictors (input nodes). From there, we would create the hidden layer (hidden nodes), followed by the prediction (output node). The neural networks would be adapted for patient data modeling.

**Phase 4:** Upon completion of patient data modeling, the predictive models would then be ready to be used in clinical psychiatric settings.

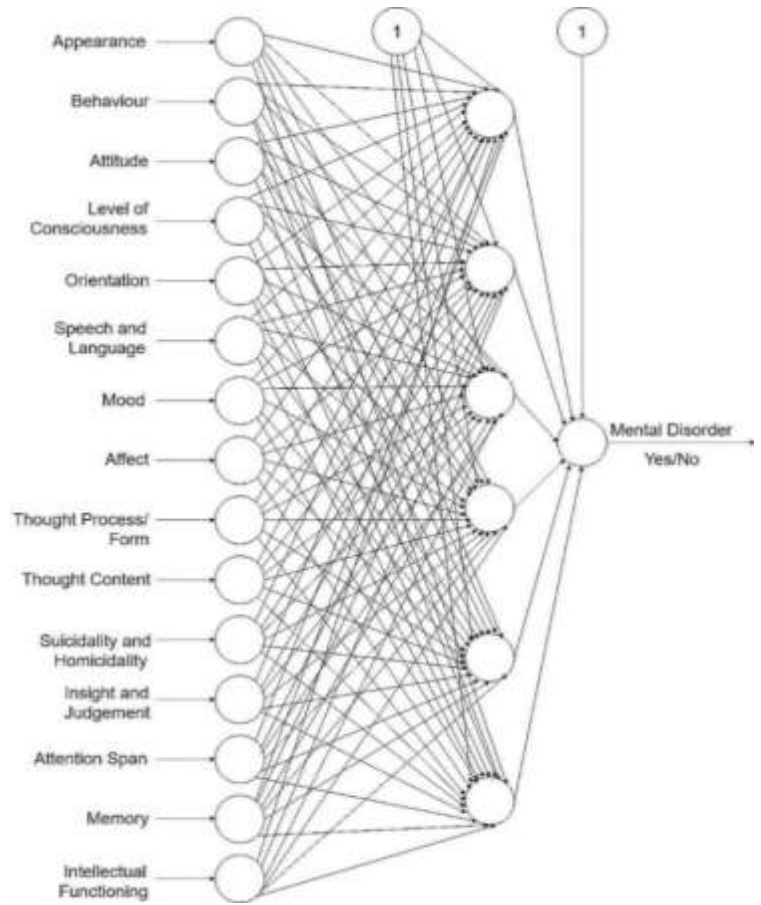

**Figure 2.** An illustration of a neural network-based model visualized to demonstrate a simplified version of a neural network used to predict mental disorders in mental health patients. The features (predictors) are listed on the left, represented by the circles which are the input nodes. The middle circles represent the hidden layer, and the individual circle on the right represents the output node.

Testing of the neural networks has not been performed and needs to be done.

*Analytics Complexity Analysis Overview*

There are two kinds of complexities that must be analyzed; the complexity of neural networks and the complexity of predictive analytics. The reason behind performing two analyses is because while some complexities of neural networks and predictive analytics may overlap, there are others that do not. Therefore, in analyzing the complexities for both, we can get a better understanding of the complexities involved in developing predictive models.

*Neural Networks Complexity Analysis*

For the complexity of neural networks, there are three kinds of complexities that could be investigated; (1) the theory of neural complexity, (2) computational complexity, and (3) information complexity. The theory of neural complexity differs from information complexity as the theory addresses the number of neurons (nodes) in the hidden layer which are necessary for the computation of a given function with assuming that full information about the function within some bias (Kon & Plaskota, 2000).

For our case, we find that the number of nodes in the hidden layer is 6 excluding the bias node. While this may not seem like a lot, this number is expected to increase due to the number of variables that will be contained within the training data that will be used to train the neural network. The more complex the information and data is (i.e. the more variables that are involved), the more complex the function will be. The expected function will therefore become something like the following:

$$E(Y) = \sum_{i=0}^{n} W_i X_i - B = Y$$

Hence, as the function becomes more complex, more nodes will be needed in the hidden layer for the computation of that function to occur. As additional variables get added into the function, the complexity of the computability of the neural network increases.

To solve the computational problem given the input distribution, a statistical query (SQ) algorithm will be required. Statistical query algorithms are algorithms that solve computational problems over an input distribution (Song, Vempala & Wilmes, 2017). As this complexity component involves showing and proving that efficient neural network training algorithms need to use stronger assumptions on the target function and input distribution, the complexity analysis for computational complexity will not be addressed here. A separate research must be conducted for properly evaluation the computational complexity of neural networks.

For information complexity, information complexity deals with information issues and numbers the examples needed to encode given tasks into neural networks (Kon & Plaskota, 2000). While we assumed that 16 predictors would be required to encode the task of predicting whether a patient has a mental disorder and what kind of mental disorder the patient has, the exact number of predictors is still up for debate. We know that diagnosing for neurodevelopmental disorders, those may require the use of neuroimaging. If brain images and brain features are added as variables into the neural networks, then numbers the examples needed to encode the task of predicting mental disorders in patients will drastically increase. As more information is needed to accurately predict the mental disorder, the more examples will be needed to encode the task into the neural network.

Combining all three complexities together, we came to the conclusion that for the complexity of the neural network for predicting mental disorders, with at least 16 predictors being involved in the neural network and more variables are likely to be added for more accurate predictions, this implies that a high numbers the examples are needed to encode the task of predicting mental disorders; that subsequently increases the number of nodes in the hidden layer that are necessary for the computation of the given function. The function itself becomes extremely intricate as a high number of nodes are needed in the hidden layer for the computation of the function.

*Predictive Analytics Complexity Analysis*

Complexity in predictive analytics is heavily tied to data complexity. The reason behind this is because predictive analytics relies on data to make predictions, which therefore implies that complexity in data would also be complexities in predictive analytics. Data complexity for our case would pertain to the size and intricacy of the patient data with the specific relevant domains being data (1) size, (2) structure, (3) variety, and (4) abstraction. For accurate predictions, a large volume of data is needed, which large data is complex. The structure of data, specifically that of relationships between variables, will be data that is intertwined with many relationships, therefore having complex data structures. From the first two stages of mental disorder treatment, we find that the patient data will likely be heterogenetic, resulting in having complex data with a diverse structure and values. Lastly, as the data will be abstracted from the patient data, this makes the data more complex than non-abstracted data.

## Limitations

The limitations associated with applying predictive analytics (and neural networks) to clinical psychiatry is more immense than expected one may expect. Already as individual fields, both suffer from their own separate kinds of limitations. Combined, not only do those limitations cumulate together, but more limitations are also added to that.

*Psychiatric Practice Limitations*

There are many limitations that exist within the practice of psychiatry. Some of the limitations that exists within the treatment of mental disorders is the availability of assessment of data, services characteristics, and challenges associated with interpreting evidence presented.

In analyzing complexities in clinical psychiatry, task and patient complexity presents numerous limitations. When it came to task complexity, the task complexity contributing factors that exist that enforce limitations onto clinical psychiatry include one or more of the following (Islam, Weil & Del Fiol, 2016):

- Unclear Goals
- Large Number of Goals
- Conflicting Goals
- Unnecessary Information
- Confusing Information
- Urgent Information
- Multiple Decision-Making Options
- Large Number of Decision Steps
- Lack of Team Coordination
- Lack of Expertise
- Decision Conflict
- Time Pressure

In addition to the task complexity contributing factors, the patient complexity contributing factors that exist that enforce even more limitations onto clinical psychiatry include one or more of the following (Islam, Weil & Del Fiol, 2016):

- Psychological Illness
- Mental Anxiety
- Significant Physical Illness
- Age
- Health Disparity
- Low Social Support
- Non-Compliant Patient

Combined, the task and patient complexity contributing factors create a difficult and challenging situation for clinician (i.e. psychiatrist) when it comes to properly perform tasks needed to effectively diagnose, assess and treat psychiatric patients successfully.

*Predictive Analytics and Neural Networks Limitations*

A limitation associated with the use of predictive analytics is that the possibility of overfitting in the predictive model can occur, where the model learns the details and noise in the training data to the extent that it negatively effects the performance of the model on new psychiatric data. This implies that the random fluctuations in the training data is picked up and learned as concepts by the predictive model. The problem is that these concepts do not apply to new psychiatric data and negatively impact the predictive model's ability to generalize. Additionally, the way predictive models cannot correct mistakes or errors made by clinicians.

Given both the task complexity and patient complexity in clinical psychiatry, mistakes and errors are bound to happen. These kinds of incorrect actions, whether it's working with inaccurate, imprecise or insufficient patient data, or working with knowing that some essential patient data is missing, cannot be corrected by predictive analytics. Regardless of whether the patient data is right or wrong, the predictive model will most likely end up making use it while learning. In addition, predictive analytics cannot address and solve many task and patient complexities, complexity factors and complexity contributing factors, and those complexities could impact the performance of predictive models and the way clinicians utilize the models;

The limitations will become most noticeable in the second and fourth phase of the phases within predictive analytics projects. Limitations in the practice of clinical psychiatry will directly impact the data that will be used by the predictive model as data collected from the sections covered in the outline of the psychiatric assessment, which that data is collected by the clinician is used to create the training data that will be used to train and develop the predictive model. This implies that any limitations in the training data can limit the performance of the machine learning technique, therefore limiting the performance of the predictive model.

As the machine learning technique that we used are neural networks, while there are several issues that could be mentioned about it, the primary and most concerning issue with neural networks is that neural networks are *black boxes*. That is, the user (i.e. clinical psychologist and psychiatrist) feeds in data and receives answers without necessarily having access to the exact decision-making process. Other limitations with neural networks is that it takes a considerable amount of time to train them, which requires an extensive amount of computing power for complex tasks and a large number of patients. What worsens the case when using predictive analytics that relies on neural networks is that as psychiatric diagnosis and treatments are extremely complex to begin with, it is guaranteed that neural networks will be required to use of significantly high amount of computing power to perform tasks pertaining to clinical psychiatry.

## Medical Ethics Issues

From the many medical ethics concerns that can be said, the two main issues that will be emphasized and discussed involves the current problems that psychiatry as a field is facing; *misdiagnosis* and *overdiagnosis* of mental disorders.

### Misdiagnoses in Psychiatry

The issue of misdiagnosis is a problem that can be traced back through the history of psychiatry. Psychiatry has and still currently faces problems when it comes to diagnosing mental disorders. It is for that reason the DSM was made; to assist clinicians with diagnosing mental disorders and treating mental health patients. In attempt to improve the accuracy of diagnoses, a total of 5 revisions to the DSM was made in total, with the DSM-5 being the most recent and the one currently being used by clinicians. Even with the DSM-5, there are problems of misdiagnoses and diagnostic errors. There are several studies that have shown high rates of misdiagnoses and diagnostic errors in daily practice. A study on diagnostic mistakes of culturally diverse individuals found that of the diverse ethnic population that they studied, close to 51% of those assessed may have been misdiagnosed as having a mild or major cognitive disorder (Daugherty et al., 2017). Another study that focused on the rate of misdiagnosis and the impact of duration of untreated illness on outcome of bipolar patients with psychotic symptoms found that 61.5% received a first diagnosis different from bipolar disorder with the DSM-5, implying that more than half of bipolar patients with psychotic symptoms receive a different diagnosis at the initial contact with psychiatric services (Altamura et al., 2015). Misdiagnoses do not only include adults, but also include children and adolescents as well. A study on misdiagnosis and missed diagnoses in foster and adopted children found that among a population of children and adolescents, the misdiagnosis rate was 80.1% for prenatal alcohol exposure (Chasnoff, Wells & King, 2015).

These are just some of the many studies on misdiagnoses that were mentioned, there are many more than can be presented. With these kinds of misdiagnoses and diagnostic errors occurring, these would heavily impact psychiatric data negatively, which would that same psychiatric data would then be used to into creating datasets for predictive analytics. As a result, when the datasets are used as training data for the neural network (or even any other machine learning technique), the predictive models that would be created would model false information. In the short run, the outcome could then lead clinicians to making more misdiagnoses and diagnostic errors rather than decrease the number of mistakes made in diagnoses. In the long run, the outcome could lead to increasing the number of biases made by clinicians based on gender, race, age, ethnicity. In both cases, these are some medical ethic issues that should be further discussed.

### Overdiagnoses in Psychiatry

Overdiagnosis of mental disorders is an emerging concern that is currently being discussed within the medical community. A study showed that of 4896 studies, 9% of the studies reported overdiagnosis of mental disorders, with the studies predominantly investigating bipolar disorder (Jenniskens et al., 2017). From 2003 to 2011, the frequency of attention deficit hyperactive disorder (ADHD) in the U. S. increased by approximately 35% (Edwards, 2016). Similarly, a recent study on overdiagnosis of ADHD in the practice of psychiatry reported prevalence rates up to 20% (Merten et al., 2017).

As neural networks can make predictions with missing variables, this allows for predictive models to be made regardless of whether all necessary patient data was collected. This could lead to an increase in diagnoses of mental disorders as patients can be served more quickly. However, with the increase of diagnoses comes an increase in treatments, which as we know, treatments that involve drug prescriptions can result in having the patient manifest undesirable side effects. To make things worse, from the conclusion made on misdiagnoses, there is a chance that a fairly large number of the diagnoses made being misdiagnoses. Thus, not only is overdiagnosis possible, but also having an increased population of people (i.e. patients) that suffer from side effects from prescribed drug and/or misdiagnoses is possible too, making it a medical ethic concern involving both the medical community and society at whole.

## Discussions

This paper was initially conducted to get a better understanding of the treatment process of mental disorder and process of predictive analytical projects, investigate how predictive analytics could be applied to clinical psychiatry, then analyze complexities, limitations and possible consequences of applying predictive analytics to clinical psychiatry. However, as the number of sources (articles, medical manuals, critical analyses, and literature reviews) decreased, the scope of the research shifted to focusing on investigating how neural networks and predictive analytics could be applied to clinical psychiatry specifically in diagnoses, while also analyzing complexities, limitations and possible consequences of applying neural networks and predictive analytics to clinical psychiatry. When conducting the literature search, there were more sources that covered neural networks, predictive analytics and psychiatry separately, but aside from few expert reviews, there were none that covered neural networks or predictive analytics, and psychiatry. As more sources covered areas outside of the predictive analytics and clinical psychiatry, the focus was therefore shifted to accommodate the existing sources.

Testing of the neural network was not possible due to the time constraint, and being unable to of obtaining data sets from medical institutions. There is an inadequate number of sources that discuss the complexities of neural networks and predictive analytics. There was one article that provided valuable information on the practice of medicine, which was useful as many components covered in the source applied to clinical psychiatry as well. However, for the complexity of neural networks and predictive analytics, there were no sources that properly covered the theory of neural complexity, computational complexity, and information complexity. From the literature search, something that can be concluded from it is that efforts are being made to understand the computability complexity of neural networks with statistical query algorithms being used to address it. Additionally, effort is currently being made to understand data complexity rather than predictive analytics. Lastly, there are several studies that investigate the situation of misdiagnoses, with few on overdiagnoses.

## Conclusions

Predictive analytics back by neural networks could potentially serve as a tool to help with the practice of psychiatry, but it is important to realize that predictive analytics has the potential to also cause harm. Misdiagnosing and overdiagnosing are two medical ethic concerns that were raised. These two were specifically selected because of the consequences they both have on both individuals and society. Only after a period and after predictive analytics has been tested and evaluated using decision curve analysis and other evaluation methods will we know whether predictive analytics can improve or worsen the field of clinical psychiatry. Just as technology can solve problems, it also can create new ones as well. No matter how perfect the predictive models and neural networks are, there are many ways and areas where things can still go wrong.

If there are recurring issues within the diagnoses and treatments of mental disorders without the involvement of predictive analytics and neural networks, there may likely be problems with their involvement. This is because neural networks learn from the training data provided by clinicians, which if there are flaws (i.e. false, imprecise, or inaccurate data) in the training data, then the neural network will pick up those flaws. The predictive model produced by neural networks may contain those flaws when making predictions and generalizations as neural networks cannot catch problematic patterns in data, which when perceived and interpreted by clinicians, those flaws may not be noticed, negatively influencing the decision of the clinician. Decisions by clinicians could then be motived by problems such as wrong assumptions, biases towards certain groups, genders, ethnicity, age group, etc. With predictive analytics possibly making diagnostics easier for clinicians, this could potentially lead to increasing the number of misdiagnoses and overdiagnoses, raising the question of whether predictive analytics and neural networks would be beneficial or detrimental to practice of clinical psychiatry, and to society.